\documentstyle[aps,prl,epsf,amsfonts]{revtex}
\begin{document}
\draft
\twocolumn[\hsize\textwidth\columnwidth\hsize\csname@twocolumnfalse%
\endcsname
\title{Boundary Effects in Chiral Polymer Hexatics}
\author{Randall D.~Kamien and Alex J.~Levine}
\address{Department of Physics and Astronomy, University of Pennsylvania,
Philadelphia, PA 19104}

\date{\today}
\maketitle
\begin{abstract}
Boundary effects in liquid-crystalline phases can be large due to
long-ranged orientational correlations.  We show that the chiral
hexatic phase can be locked into an apparent three-dimensional
N+6 phase via such effects.  Simple numerical estimates suggest
that the recently discovered ``polymer hexatic'' may actually be
this locked phase.
\end{abstract}
]

Liquid crystals provide a tabletop laboratory to study broken symmetries and
the resulting low-energy Nambu-Goldstone modes \cite{CL}.  Viewing
two-dimensional
crystallization in this way led to the proposal of an intervening
phase between the liquid and the solid: the hexatic \cite{HN},
a two-dimensional phase with long-range orientational order but short-range
positional order.  In three dimensions, arguments based on Landau theory
suggest that there is no similar orientationally ordered phase
between the three-dimensional liquid and solid \cite{LL}.  However, between
a nematic phase and a crystal phase, the mesogens can have
hexatic order in the plane perpendicular to the nematic director \cite{TONER}.
While
this ``N+6'' phase could, in principle, exist (it is not terribly different
from a biaxial-nematic phase) it was not immediately discovered.  Recently,
however, Strey, {\sl et al.} \cite{STREY} have examined a phase
of DNA, having a structure consistent with N+6 order, which they call the
polymer hexatic.  In this letter, we will consider the effects of boundaries
on an N+6 phase consisting of chiral molecules, such as DNA.  We will argue
that while symmetry predicts that the hexatic bond order should twist
\cite{LCBO}
surface effects can ``lock-in'' a preferred set of directions which can
effectively unwind the twisting hexatic.

\begin{figure}
\epsfxsize=3.0truein
\vskip15pt
\centerline{\epsfbox{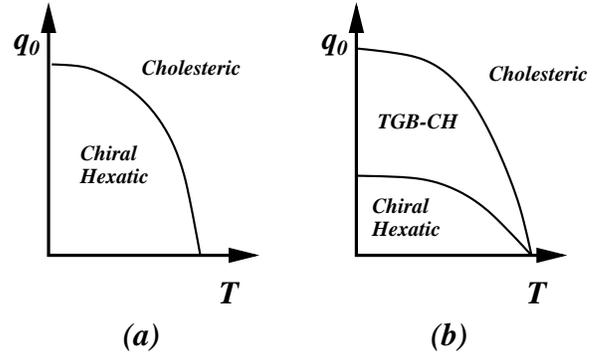}}
\vskip15pt
\caption{Phase diagram of a chiral mesogen that can form an N+6 phase.
In the cholesteric phase there is no hexatic order ($\psi_6=0$) and
$\bbox{\nabla}\times{\bf n}\ne 0$.  In the chiral-hexatic phase
$\bbox{\nabla}\times{\bf n}=0$ but $\psi_6\ne 0$ and $\bbox{\nabla}\theta_6
=\tilde q_0 {\bf n}$.
(a) Type-I behavior with no intervening defect phase. (b) Type-II behavior
in which a twist-grain-boundary phase (akin to the Abrikosov
flux-phase in a superconductor) intervenes between the cholesteric
and chiral-hexatic phase.}
\label{f2}
\end{figure}

The phase sequence of a chiral mesogen usually starts with the cholesteric
at the highest temperatures or lowest concentrations.  As the temperature
is lowered or the density increased, the mesogens can form any number of
phases,
including the (achiral) smectic-A phase, the smectic twist-grain-boundary (TGB)
phase
\cite{RL} or the chiral hexatic \cite{LCBO}.  In the last case, the free
energy which governs the transition from cholesteric to chiral hexatic depends
on both the nematic director ${\bf n}$ and the complex hexatic order parameter
$\psi_6$:
\begin{eqnarray}
F_{\rm bulk}&=\int d^3\!x\,\Bigg\{&\left\vert\left(\bbox{\nabla}
-i\tilde q_0{\bf
n}\right)\psi_6\right\vert^2
+ r\left\vert\psi_6\right\vert^2 + u\left\vert\psi_6\right\vert^4\nonumber\\
&&+{1\over 2}K_1\left(\bbox{\nabla}\cdot{\bf n}\right)^2 +
{1\over 2}K_2\left({\bf n}\cdot\bbox{\nabla}\times{\bf n} +
q_0\right)^2\nonumber\\
&&+{1\over 2}K_3\left[{\bf n}\times\left(\bbox{\nabla}\times{\bf
n}\right)\right]^2\Bigg\}
\label{e1}
\end{eqnarray}
where $r$ and $u$ are Landau parameters and $K_i$ are the Frank elastic
constants.  
In the cholesteric phase, $\langle\,\psi_6\,\rangle=0$ and (\ref{e1}) reduces
to the free energy of a cholesteric with equilibrium pitch $P=2\pi/q_0$.  When
hexatic order persists $\psi_6=\vert\psi_6\vert e^{i6\theta_6}$ and, as
in the Meissner phase of superconductors \cite{RL},
the nematic director satisfies
$\bbox{\nabla}\times{\bf n}=0$.  However, in this phase, it is straightforward
to see that the hexatic bond-order rotates about the average nematic direction:
$\bbox{\nabla}\theta_6 = \tilde{q}_0{\bf n}_0$.  Recalling the phenomenology
of the superconductor or smectic-A liquid crystal \cite{DG} we
emphasize that if the director unwinds then the hexatic bond-order
{\sl must} twist.  Further, we expect the cholesteric
pitch to unwind continuously to infinity at the cholesteric--to--chiral-hexatic
phase transition, as has, in fact, been observed \cite{STREY}.  
Possible phase diagrams are shown in Figure \ref{f2}.

\begin{figure}
\epsfxsize=2.6truein
\vskip15pt
\centerline{\epsfbox{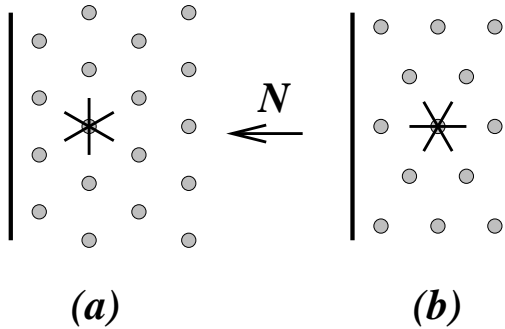}}
\vskip15pt
\caption{Hexagonally close-packed disks near a wall with surface
normal ${\bf N}$.  Each hexagonal
crystallite defines a set of hexatic axes shown as six-pointed arrows.
In (a) we show a favorable packing for ``surface loving'' disks,
while in (b) we show a packing of ``surface hating'' disks.  For either
case, deviations away from this packing are disfavored entropically.}
\label{f1}
\end{figure}

Deep in the chiral-hexatic phase the nematic order freezes \cite{LCBO} and
we can rewrite (\ref{e1}) in the ``London limit'':
\begin{equation}
F_{\rm bulk}=
\int d^3\!x  \Bigg\{{1\over 2}K_A\left(\bbox{\nabla}\theta_6\right)^2
-K_A\tilde q_0{\bf n}_0\cdot\bbox{\nabla}\theta_6 \Bigg\} ,
\label{e2}
\end{equation}
where $K_A=72\vert\psi_6\vert^2$.
In the following we take the average
nematic axis to be ${\bf n}_0={\bf z}$.
This free energy, however, applies only deep inside the sample.  In general
there will be surface terms which can bias the hexatic bond-order direction.
For instance, if the boundary is a flat surface, then the mesogens
will either be ``surface loving'' or ``surface hating'' and will try to
maximize or minimize their contact with the surface, respectively.
If the surface tension is large, the first few layers of molecules will
be locked and then simple geometric
arguments, akin to the classic Onsager treatment of nematics \cite{DGP},
show that there is an entropic penalty for the hexatic order
to rotate away from the preferred direction, as shown in Figure \ref{f1}.

To model the effect of this surface interaction, we add to (\ref{e2})
the surface term:
\begin{equation}
F_{\rm surface} = \int_{\partial\Omega} dsdz\, h\cos(6\theta_6)
\end{equation}
where $\Omega$ is the volume of the sample and $\partial\Omega$ is the boundary
in the $xy$-plane.  The sign of $h$ is  determined by
the surface affinity of the mesogens, while the magnitude of $h$ can
be estimated through an excluded-volume, entropic analysis.
Taking the mesogens to be polymers (so that in Figure \ref{f1}, the disks
are the polymer cross-sections in the $xy$-plane), we consider
an hexagonal crystallite of extent $\xi$ in the $xy$-plane and length $L_P$
along $\hat{\bf z}$, rotating away from its preferred direction.  Presumably
$L_P$ is precisely the polymer persistence length, which is the
typical correlation of the polymer positions along the nematic axis.
In this case, the entropy cost for a rotation $\theta$ is
\begin{eqnarray}
\Delta F&=&-k_{\scriptscriptstyle B}T\Delta S = -k_{\scriptscriptstyle B}T
N\ln\left\{{L_P\xi^2\left[1-\gamma\cos(6\theta)\right]\over
L_P\xi^2}\right\}\nonumber\\
&\approx& k_{\scriptscriptstyle B}TN\gamma\cos(6\theta)
\label{e4}
\end{eqnarray}
where $\gamma<1$ is a factor of order unity, $N=V/(L_P\xi^2)$ is
the number of polymer crystallites and $V$ is the volume of the system.
Converting (\ref{e4}) into an integral over the surface gives
$\vert h\vert \approx k_{\scriptscriptstyle B}T/(L_P\xi)$.  For
the polymer hexatic phase studied in \cite{STREY}, $L_P\approx 500\AA$
and $\xi\approx 400\AA$.

We propose that this surface anchoring can prevent the equilibrium
twisting that a chiral-hexatic phase must exhibit whenever the nematic
director is aligned.  To pursue this, we consider configurations
in which the hexatic bond-order is uniform at any constant height $z$.
This should be reasonable deep inside the hexatic phase where there
is true long-range bond order.  The free energy
is:
\begin{equation}
F_{\rm eff} = \int dz \left\{{K\over 2}\left(\partial_z\theta_6\right)^2
- K\tilde q_0\partial_z\theta_6 + H\cos\left(6\theta_6\right)\right\},
\label{e8}
\end{equation}
where $K\equiv K_A A$, $H\equiv hL$, $A$ is the cross-sectional
area of the sample and $L$ is the length of the sample along the
wall which provides the
anchoring.  This effective free energy has frequently been studied before
in the context of commensurate-incommensurate transitions.
For small values of $\tilde q_0$ the $\cos(6\theta_6)$ term prevents
{\sl any twisting}, while for larger values the system admits
soliton solutions which let $\theta_6$ slip by $2\pi/6$.
In equilibrium, it is straightforward \cite{PNT} to show that
a soliton is the lowest energy configuration when
\begin{equation}
\vert\tilde q_0\vert \ge {4\over \pi}\sqrt{{L\over A}{h\over K_A}}.
\label{yesorno}
\end{equation}
In addition to the estimate for $h$, we can estimate $K_A$
via dimensional analysis: $K_A\sim k_{\scriptscriptstyle B}T/\xi$.  How
should we estimate $\tilde q_0$?  In the cholesteric phase of DNA, the
pitch is on the order of $1 \mu m$.  We can use this to estimate the
rate at which a polymer bundle twists about the average nematic axis.
To do so we consider a bundle of $N$ polymers, with separation $a$.  We
let them twist about the center of the bundle so that the polymer
that crosses $z=0$ at $(\rho,\phi)$ has conformation \cite{ILCC}
\begin{equation}
\label{twisted}
{\bf R}(z; \rho, \phi ) =
\left[\rho \cos(\tilde q_0 z + \phi ), \rho \sin(\tilde q_0 z + \phi), z\right]
\end{equation}
We calculate the energy for a twisted bundle of radius $\rho_0$ and
length $L_z$.  There
are two contributions to the energy: the bending energy of the individual
polymers
and the Frank free energy of the liquid crystal deformation.  The first energy
is proportional to the bending modulus $\kappa$.  If the polymers sit at
radii $\rho_i$ and have an average spacing $a$, then, for $\tilde q_0\rho\ll 1$
we have:
\begin{eqnarray}
\label{bend}
F_{\rm bend}&=&\sum_i {\kappa\over 2}\int dz\,
\left\vert{d^2{\bf R}\over dz^2}\right\vert^2
\left[1+\left(\rho_i\tilde q_0\right)^2\right]^{-3/2}\nonumber\\
&\approx& {\kappa\pi\over a^2}\int dz\int_0^{\rho_0}\rho d\rho
\rho^2\tilde q_0^4\nonumber\\
&\approx& {L_z\kappa\pi\over 4a^2}\rho_0^4\tilde q_0^4 = {L_z\kappa\pi
N^2a^2\tilde q_0^4\over 4}
\end{eqnarray}
where $N$ is the number of polymers in the bundle.  The liquid crystal
deformation
has no splay and we have just accounted for the bend deformations.  The
remaining
energetic contribution is from the twist:
\begin{equation}
\label{twist}
\frac{F_{\rm twist}}{L_z}
= {NK_2\pi a^2\over 2} \left( 2 \tilde{q}_0 + q_0 \right)^2.
\end{equation}
Finally, minimizing the total free energy $F=F_{\rm bend} + F_{\rm twist}$ over
$\tilde q_0$,
we find:
\begin{equation}
\left[1 +\frac{N\kappa\tilde q_0^2}{4 K_2}\right]\tilde q_0 = -\frac{ q_0}{2}
\label{balance}
\end{equation}
The bending modulus is $\kappa =k_{\scriptscriptstyle B}TL_P$ where $L_P$ is
the persistence length,
while $K_2 = k_{\scriptscriptstyle B}T/d$ where $d$ is the Odijk length, given
by $d\sim (L_P a^2)^{1/3}$.
Thus
an estimate of the coefficient of $\tilde q_0^2$ is
$N \pi^2 L_P^{4/3}a^{2/3}/P^2$ where
$P$ is the hexatic pitch.  The
experiment \cite{STREY} finds a translational correlation length on the order
of
ten $a=4 nm$ intermolecular separations so
presumably each twisting bundle is composed of
$N\sim 100$ molecules.  In this case,
$N\kappa /K_2 \approx 10^{-2} \mu {\rm m}^2$.  If we take a typical range
$1 \mu \mbox{m} < 2 \pi /q_0 <
10 \mu \mbox{m}$ then (\ref{balance}) shows that
$2 \mu \mbox{m} < 2 \pi /\tilde{q}_0 < 20 \mu \mbox{m}$.
This follows from the fact that
though the persistence length is long for DNA, it is still two orders of
magnitude smaller than the cholesteric pitch: twisting on this length scale
is easy for the polymers.  .

Using
(\ref{yesorno}), we find that twisting (through the introduction of
solitons in the $\theta_6$ field) becomes unfavorable for systems of linear
dimension smaller than a certain critical length.  For system sizes such that
$L < 16/\left( L_P \pi^2 \tilde{q}_0^2 \right) $
the boundary terms dominate the free energy so the bond order
is effectively pinned by these surface effects.  From the above estimates of
the
equilibrium pitch of the hexatic bond order we find that the
largest system
which allows surface pinning is on the order of $300 \mu{\rm m}$ when the 
pitch is $20 \mu\rm{m}$.
The sample was on the order of $1 {\rm mm}$ \cite{STREYPRI}
so the outer boundaries of the sample are possibly  too far away to cause
lock-in.  However, it is possible that sample preparation
introduced interior walls.
A wall spacing on the order of $\sim 300\mu{\rm m}$
would imply internal regions of roughly $10^7$ DNA molecules.  If, on the
other hand, we were to take the smallest reasonable pitch of $1 \mu{\rm m}$, we
would find
a wall spacing of $3 \mu{\rm m}$, which is {\sl unreasonable}.  Indeed, it is
likely that in the hexatic phase, intermolecular correlations can modify the
chiral
interaction as argued in \cite{STREY} and \cite{HKL} and lead to a lengthening
of
the pitch.  It is thus reasonable that in the hexatic phase the chiral strength
can be reduced and we are somewhere between these two extremes.

The data, if taken to represent
only the bulk physics, shows that $2 \pi /\tilde{q}_0 \sim 1 {\rm mm}$
since the hexatic
order does not twist by $2\pi/6$ inside the illuminated region.
However, with walls one millimeter away, the data is consistent
with an equilibrium pitch of $44 \mu{\rm m}$, not that far from our
na\"\i ve estimate of the pitch.  Thus, bearing in mind the Landau
free energy of this system, one can only conclude that the hexatic
pitch is longer than $44 \mu{\rm m}$, but not {\sl infinite}!
The absence of the observation of a finite
hexatic pitch can be attributed to the surface pinning effects. If
we take our estimate of the hexatic pitch and make the reasonable
assumption of a few internal walls in the sample, or if
we imagine the hexatic pitch to be within a factor of two of typical
pitches, it is clear that
such a pinned state is expected.
In either
case the effects of surface pinning are required to interpret the data.
As further evidence for our pinning hypothesis, it is known that
the hexatic director tends to align with the long axis of the sample
\cite{STREYPRI}.

\begin{figure}
\epsfxsize=3.0truein
\vskip15pt
\centerline{\epsfbox{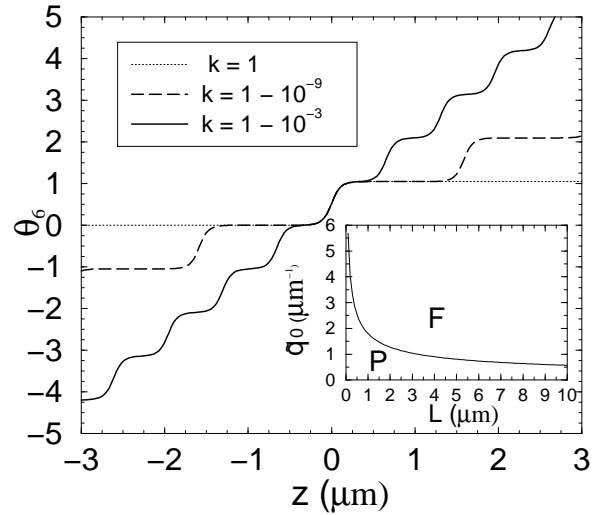}}
\vskip15pt
\caption{%
Three solutions for the bond-order field as a function of height.
We show the one soliton ($k=1$, the dotted line) solution, as well
as multi-soliton solutions for $k<1$.
The parameter $k$ is determined by the sample dimension $L$
and the equilibrium pitch
$\tilde{q}_0$ allowing
the construction of the phase diagram shown in the inset.
In the``pinned'' phase
(P) the bond-order field $\theta_6$ is prevented from twisting by the
surface pinning while in the ``free'' (F) phase the bond-order can twist.  Upon
approaching the phase boundary (solid line) from below,
the bond order begins to twist by the creation of solitons.  The actual
phase boundary line is chosen by arbitrarily assigning a maximum (non-zero)
soliton density as described in the text.%
}
\label{soliton_fig}
\end{figure}
Since both sides of (\ref{yesorno})
are the same order of magnitude the condition to prevent solitons
is, at best, marginally satisfied.
What happens if we have solitons?  We shall see that even close to
the transition a state with solitons will still have an achiral hexatic
signature.
When (\ref{yesorno}) is an equality, it becomes energetically
favorable for
the system to introduce one soliton into the bond order field,
$\theta_6(z)$.  Such a field configuration, in which the hexatic bond order
jumps by $2 \pi/6$ near $z=0$, is shown by the dotted line in
Figure \ref{soliton_fig}.  At low densities two solitons interact via a weak
exponential 
repulsion with a range set by 
the distance over which $\theta_6(z)$ changes rapidly with $z$.

This weak interaction of two solitons leads to the rapid
proliferation
of such twists in the system upon tuning the system parameters, $L$ and
$\tilde{
q}_0$.  Once these parameters are tuned so that the chemical potential of
a soliton becomes negative, numerous solitons are
spontaneously created until their weak mutual repulsion returns the soliton
chemical potential to zero.  In this case the bond-order field takes the
form \cite{Frank:49}:
\begin{equation}
\label{theta-eqn}
\theta_6(z) = \frac{ \pi }{6} \left[ 1 + \frac{ 2}{\pi } \mbox{am}
\left( \frac{ 6 z}{k} \sqrt{\frac{ H}{K_A}}, k \right) \right]
\end{equation}
where am$(x,m)$ is the inverse of the incomplete elliptic
integral of the second kind and the parameter $k$
is related to the energy via $E(k^2)=k\pi\tilde q_0\sqrt{L_PL}/4$ where
$E(x)$ is the complete elliptic integral of the second kind.  The solution
to (\ref{theta-eqn})
describes a regular array
of solitons with a density controlled by the
value of the parameter $k$.

As shown in Figure \ref{soliton_fig} the parameter $k$ controls the
equilibrium density of solitons in the system.  When $k=1$, there is precisely
one
soliton in the infinite system and, as $k$ decreases from unity, the soliton
density grows monotonically.  When $k=0.95$ this soliton density
reaches a point where mutually overlapping solitons join to form an
almost uniform gradient in the bond order.
The system returns to
its ``free'' behavior with $\theta_6 \left( z \right)  \propto z$.  On the
other hand, if $k$ is larger than unity, there are no solitons in the system
in equilibrium.  The bond-order field
is completely pinned by the edges of the sample so that
$ \theta_6 \left( z \right)$ is constant.

The experimental signature of the pinned phase is a six-fold modulation of the
in-plane scattering intensity.  It can, in fact, 
persist into regions of the phase diagram
which allow modest densities of solitons.
The six-fold modulation of the scattering intensity should
remain evident as long as the density of solitons is small enough so
that distinguishable
plateaus in the $\theta_6$ shown in Figure \ref{soliton_fig} exist. We take
a somewhat arbitrary but nevertheless conservative estimate of the
transition to be the point at which the plateaus are twice as long as the
intervening regions where the $\theta_6$ field changes rapidly \cite{plat}.
By this criterion the solid  curve in Figure \ref{soliton_fig}
represents a free phase configuration of $\theta_6$.
The division of the parameter space between the pinned and freely-twisting 
bond order selects a critical value of
$k = k^\star$.
Solving for $k^\star$ leads to a phase boundary between
pinned and free states of the form: $ \tilde{q}_0 L^{1/2} =\mbox{const}$.
This curve
(with the constant set by $k^\star$) is shown in the inset of
Figure \ref{soliton_fig}.
Additionally, the soliton lattice could, in principle, be observed via scattering
due to its periodic structure along $\hat z$.

Finally, we comment on the role of the thermal effects.
Fluctuation effects become
important near the boundary of
the pinned and free phases
since there the system exhibits a delicate balance of the bulk elastic
and surface
energies.
The qualitative effect of thermally produced
solitons is to
move the phase boundary down and to the left in the inset of
Figure \ref{soliton_fig}.
To assess
the magnitude of this shift,
we compare the surface free energy cost
to $k_{\rm B} T$.
In a system with a size as small as $L \sim 10 \mu {\rm m}$ the surface energy
is on the order of $ 10^3 k_{\rm B} T$.
The transition
is controlled then by the balance of energies three orders of magnitude greater
than the thermal energy, so we conclude that thermal
effects are negligible
except in the narrow region along the phase
boundary where
the surface and bulk energies are balanced to within $0.1\%$.
Additionally, at higher temperatures the long-range hexatic order will
melt, leaving a surface hexatic along with a bulk cholesteric.
The details of this transition could be confirmed through further
experiments along the lines of \cite{STREY}.

It is a pleasure to acknowledge stimulating conversations with T.C.~Lubensky,
V.A.~Parsegian, R.~Podgornik and H.~Strey.  RDK was supported by
NSF CAREER Grant DMR97-32963, an award from Research Corporation and a gift
from L.J.~Bernstein.  AJL was supported in part by the
Donors of The Petroleum Research Fund,
administered by the American Chemical Society.


\begin{references}
\bibitem{plat}%
We define the plateau to be the set of points (z) for which $\theta_6(z)$
remains within 5\% of $2 n \pi /6$ with $n$ an integer.
\end{references}
\end{document}